\documentclass[runningheads]{llncs}
\usepackage{graphicx}
\usepackage{float}
\usepackage[section]{placeins}
\usepackage{url}

\begin{document}
\title{What Would You Ask the Machine Learning Model? Identification of User Needs for Model Explanations Based on Human-Model Conversations.}
\titlerunning{What Would You Ask the Machine Learning Model?}
%
\author{Micha{\l} Ku\'zba\inst{1, 2}\orcidID{0000-0002-9181-0126} \and
Przemys{\l}aw Biecek\inst{1, 2}\orcidID{0000-0001-8423-1823} }
\authorrunning{M. Kuzba, P. Biecek}
%
\institute{Faculty of Mathematics, Informatics and Mechanics, University of Warsaw, Poland
\and
Faculty of Mathematics and Information Science, Warsaw University of Technology, Poland \\
\email{kuzba.michal@gmail.com}}
\maketitle              
\begin{abstract}
Recently we see a rising number of methods in the field of eXplainable Artificial Intelligence. To our surprise, their development is driven by model developers rather than a study of needs for human end users. The analysis of needs, if done, takes the form of an A/B test rather than a study of open questions. To answer the question ``What would a human operator like to ask the ML model?'' we propose a conversational system explaining decisions of the predictive model. In this experiment, we developed a chatbot called \verb'dr_ant' to talk about machine learning model trained to predict survival odds on Titanic. People can talk with \verb'dr_ant' about different aspects of the model to understand the rationale behind its predictions. Having collected a corpus of 1000+ dialogues, we analyse the most common types of questions that users would like to ask.
To our knowledge, it is the first study which uses a conversational system to collect the needs of human operators from the interactive and iterative dialogue explorations of a predictive model.

\end{abstract}

\section{Introduction}

Machine Learning models are widely adopted in all areas of human life. As they often become critical parts of the automated systems, there is an increasing need for understanding their decisions and ability to interact with such systems. Hence, we are currently seeing the growth of the area of eXplainable Artificial Intelligence (XAI). For instance, Scantamburlo et al. \cite{HAIconseq} raise an issue of understanding machine decisions and their consequences on the example of computer-made decisions in criminal justice. This example touches upon such features as fairness, equality, transparency and accountability.
Ribera \& Lapedriza \cite{Ribera2019CanWD} identify the following motivations for why to design and use explanations:  system verification, including bias detection;  improvement of the system (debugging);  learning from the system's distilled knowledge;  compliance with legislation, e.g. ``Right to explanation'' set by EU;  inform people affected by AI decisions.

We see the rising number of explanation methods, such as LIME \cite{lime} and SHAP \cite{SHAP} and XAI frameworks such as AIX360 \cite{arya2019explanation}, InterpretML \cite{nori2019interpretml}, DALEX \cite{DALEX}, modelStudio \cite{modelStudio}, exBERT \cite{hoover2019exbert} and many others.
These systems require a systematic quality evaluation \cite{ExplainingXAI,HAIsystems,XAievaluation}.
For instance, Tan et al. \cite{WhyTrust} describe the uncertainty of explanations and Molnar et al. 
\cite{XAIquant} describe a way to quantify the interpretability of the model.

These methods and toolboxes are focused on the model developer perspective. Most popular methods like Partial Dependence Plots, LIME or SHAP are tools for a post-hoc model diagnostic rather than tools linked with the needs of end users. But it is important to design an explanation system for its addressee (explainee). Both form and content of the system should be adjusted to the end user. And while explainees might not have the AI expertise, explanations are often constructed by engineers and researchers for themselves \cite{DBLP:journals/corr/abs-1712-00547}, therefore limiting its usefulness for the other audience \cite{DBLP:journals/corr/abs-1806-08055}.

Also, both the form and the content of the explanations should differ depending on the explainee's background and role in the model lifecycle.
Ribera \& Lapedriza \cite{Ribera2019CanWD} describe three types of explainees: AI researchers and developers, domain experts and the lay audience.
Tomsett et al. \cite{tomsett2018interpretable} introduce six groups: creators, operators, executors, decision-subjects, data-subjects and examiners.
These roles are positioned differently in the pipeline. Users differ in the background and the goal of using the explanation system. They vary in the technical skills and the language they use. Finally, explanations should have a comprehensible form -- textual, visual or multimodal.
Explanation is a cognitive process and a social interaction \cite{Assady2019}. Moreover, interactive exploration of the model allows to personalize the explanations presented to the explainee \cite{Sokol_2020}.

Arya et al. identify a space for interactive explanations in a tree-shaped taxonomy of XAI techniques \cite{arya2019explanation}. However, AIX360 framework presented in this paper implements only static explanations. Similarly, most of the other toolkits and methods focus entirely on the static branch of the explanations taxonomy.
Sokol \& Flach \cite{ijcai2018-836} propose conversation using class-contrastive counterfactual
statements.
This idea is implemented as a conversational system for the credit score system’s
lay audience \cite{ijcai2018-865}.
Pecune et al. describe conversational movie recommendation agent explaining its recommendations \cite{recommenders}.
A rule-based, interactive and conversational agent for explainable AI is also proposed by Werner \cite{Werner2020ExplainableAT}.
Madumal et al. propose an interaction protocol and identify components of an explanation dialogue \cite{Madumal2019AGI}.
Finally, Miller \cite{miller2017explanation} claims that truly explainable agents will use interactivity and communication.

\enlargethispage{\baselineskip} 

To address these problems we create an open-ended dialog based explanation system. We develop a chatbot allowing the explainee to interact with a predictive model and its explanations. We implement this particular system for the random forest model trained on Titanic dataset \cite{titanicData,ema}. However, any model trained on this dataset can be plugged into this system. Also, this approach can be applied successfully to other datasets and much of the components can be reused.

Our goal is twofold. Firstly, we create a working prototype of a conversational system for XAI.
Secondly, we want to discover what questions people ask to understand the model. This exploration is enabled by the open-ended nature of the chatbot. It means that the user might ask any question even if the system is unable to give a satisfying answer for each of them.

There are engineering challenges of building a dialogue agent and the ``Wizard of Oz'' proxy approach might be used as an alternative \cite{Sokol_2020,Jentzsch}. In this work however, we decide to build such a system. With this approach we obtain a working prototype and a scalable dialogue collection process.

As a result, we gain a better understanding of how to answer the explanatory needs of a human operator. With this knowledge, we will be able to create explanation systems tailored to explainee's needs by addressing their questions. It is in contrast to developing new methods blindly or according to the judgement of their developers.

We outline the scope and capabilities of a dialogue agent (Section~\ref{dialogue_system}). In Section~\ref{architecture_section}, we illustrate the architecture of the entire system and describe each of the components. We also demonstrate the agent's work on the examples.  Finally, in Section~\ref{results_section}, we describe the experiment and analyze the collected dialogues.

\section{Dialogue system}
\label{dialogue_system}

This dialogue system is a multi-turn chatbot with the user initiative. It offers a conversation about the underlying random forest model trained on the well-known Titanic dataset. We deliberately select a black box model with no direct interpretation together with a dataset and a problem that can be easily imagined for a wider audience. 
The dialogue system was built to understand and respond to several groups of queries:
\begin{itemize}
    \item \textbf{Supplying data} about the passenger, e.g. specifying age or gender. This step might be omitted by impersonating one of two predefined passengers with different model predictions.
    \item \textbf{Inference} -- telling users what are their chances of survival. Model imputes missing variables.
    \item \textbf{Visual explanations} from the Explanatory Model Analysis toolbox \cite{ema}: Ceteris Paribus profiles \cite{pyCeterisParibus} (addressing ``what-if'' questions) and Break Down plots \cite{2019arXiv190311420G} (presenting feature contributions). Note this is to offer a warm start into the system by answering some of the anticipated queries. However, the principal purpose is to explore what other types of questions might be asked.
    \item \textbf{Dialogue support} queries, such as listing and describing available variables or restarting the conversation.
\end{itemize}

This system was firstly trained with an initial set of training sentences and intents. After the deployment of the chatbot, it was iteratively retrained based on the collected conversations. Those were used in two ways: 1) to add new intents, 2) to extend the training set with the actual user queries, especially those which were misclassified. The final version of the dialogue agent which is used in the experiment at Section~\ref{results_section} consists of 40 intents and 874 training sentences.

\section{Implementation}
\label{architecture_section}
\begin{figure*}[ht]
    \centering
    \includegraphics[width=1\linewidth]{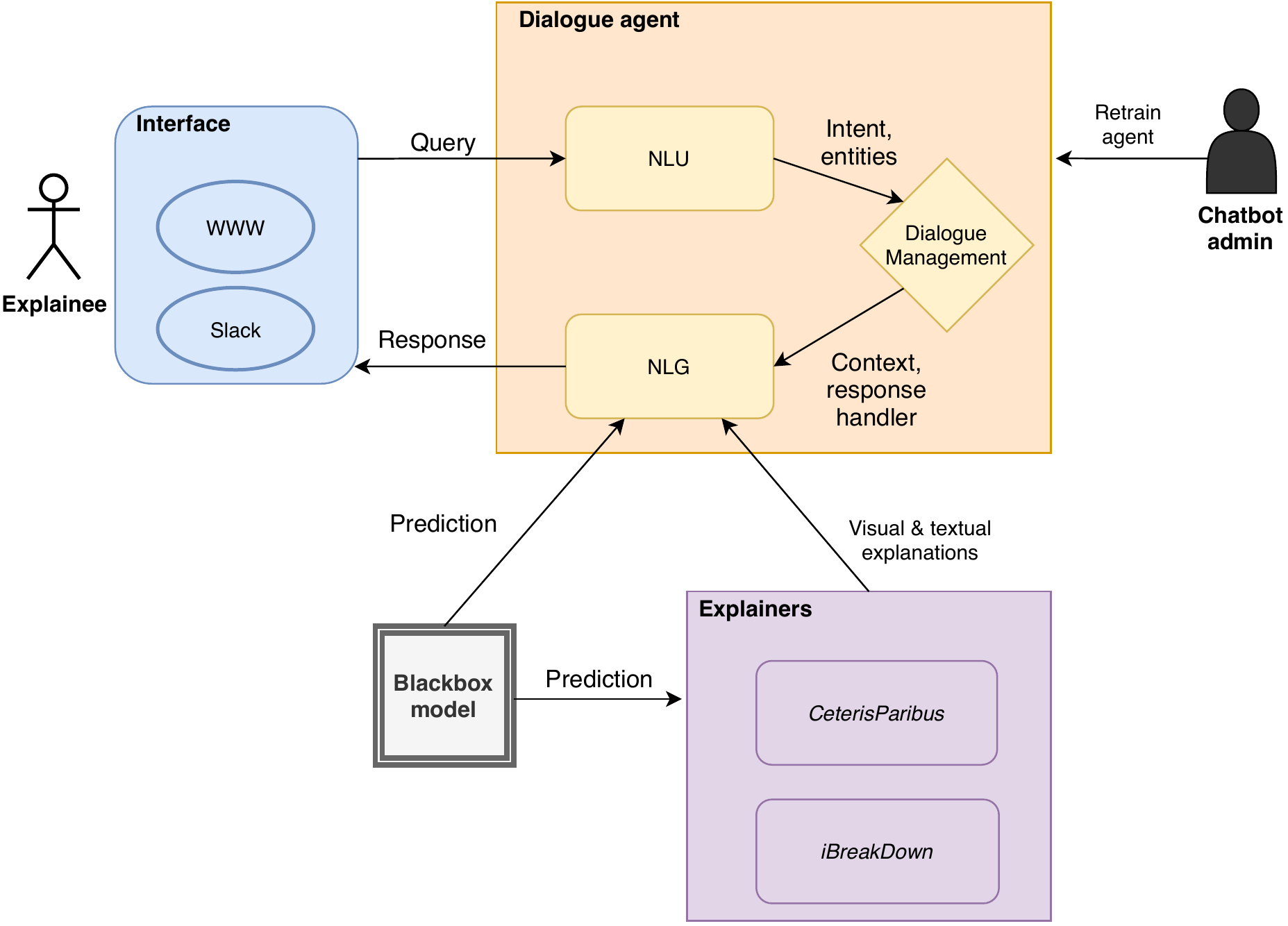}
    \caption{Overview of the system architecture. \textbf{Explainee} uses the system to talk about the \textbf{blackbox model}. They interact with the system using one of the \textbf{interfaces}. The conversation is managed by the \textbf{dialogue agent} which is created and trained by the \textbf{chatbot admin}. To create a response system queries the \textbf{blackbox model} for its predictions and \textbf{explainers} for visual explanations. }
    \label{architecture}
\end{figure*}

A top-level chatbot architecture is depicted in Figure~\ref{architecture}.
The system consists of several components:
\begin{enumerate}

    \item \textbf{Explainee} \\
    Human operator -- addressee of the system. They chat about the blackbox model and its predictions. 
    
    \item \textbf{Interface} \\
    This dialogue agent might be deployed to various conversational platforms independently of the backend and each other. The only exception to that is rendering some of the graphical, rich messages. We used a custom web integration as a major surface. It communicates with the dialogue agent's engine sending requests with user queries and receiving text and graphical content.     
    The frontend of the chatbot uses \texttt{Vue.js} and is based on \texttt{dialogflow}\footnote{\url{https://github.com/mishushakov/dialogflow-web-v2}} repository.
    It provides a chat interface and renders rich messages, such as plots and suggestion buttons.
    This integration allows to have a voice conversation using the browser's speech recognition and speech synthesis capabilities.
    
    \item \textbf{Dialogue agent} \\
    Chatbot's engine implemented using \texttt{Dialogflow} framework and \texttt{Node.js} fulfilment code run on \texttt{Google Cloud Functions}.

    \begin{itemize}
        \item \textbf{Natural Language Understanding (NLU)} \\
        The Natural Language Understanding component classifies query intent and extracts entities. This classifier uses the framework's builtin rule-based and Machine Learning algorithms.
        NLU module recognizes 40 intents such as posing a what-if question, asking about a variable or specifying its value. It was trained on 874 training sentences. Some of these sentences come from the initial subset of the collected conversations. Additionally, NLU module comes with 4 entities -- one for capturing the name of the variable and 3 to extract values of the categorical variables -- gender, class and the place of embarkment. For numerical features, a builtin numerical entity is utilized. See examples in Section~\ref{nlu_examples}.

        \item \textbf{Dialogue management} \\
        It implements the state and context. Former is used to store the passenger's data and the latter to condition response on more than the last query. For example, when the user sends a query with a number it might be classified as age or fare specification depending on the current context.
        
        \item \textbf{Natural-language generation (NLG)} \\
        Response generation system. To build a chatbot's utterance the dialogue agent might need to use the explanations or the predictions. For this, the NLG component will query explainers or the model correspondingly. Plots, images and suggestion buttons which are part of the chatbot response are rendered as rich messages on the front end.
    \end{itemize}
    
    \item \textbf{Blackbox model} \\
    A random forest model was trained to predict the chance of survival on Titanic\footnote{You can download the model from the \path{archivist} \cite{archivist} database  with a following hook: \url{archivist::aread("pbiecek/models/42d51")}.}. The model was trained in R \cite{Rcran} and converted into REST api with the \verb'plumber' package \cite{plumber}. The random forest model was trained with default hyperparameters. Data preprocessing includes imputation of missing values. The performance of the model on the test dataset was AUC 0.84 and F1 score 0.73.

    \item \textbf{Explainers} \\
    REST API exposing visual and textual model explanations from \verb'iBreakDown' \cite{2019arXiv190311420G} and \verb'CeterisParibus' \cite{pyCeterisParibus} libraries. They explore the blackbox model to create an explanation. See the \verb'xai2cloud' package \cite{xai2cloud} for more details.
    
    \item \textbf{Chatbot admin} \\
    Human operator -- developer of the system. They can manually retrain the system based on misclassified intents and misextracted entities. For instance, this dialogue agent was iteratively retrained based on the initial subset of the collected dialogues.
    
\end{enumerate}
This architecture works for any predictive model and tabular data. Its components differ in how they can be transferred for other tasks and datasets\footnote{The source code is available at \url{https://github.com/ModelOriented/xaibot}.}. The user interface is independent of the rest of the system. When a dataset is fixed, the model is interchangeable. However, the dialogue agent is handcrafted and depends on the dataset as well as explainers. Change in a dataset needs to be at least reflected in an update of the data-specific entities and intents. For instance, a new set of variables needs to be covered. It is also followed by modifying the training sentences for the NLU module and perhaps some changes in the generated utterances.  Adding a new explainer might require adding a new intent. Usually, we want to capture the user queries, that can be addressed with a new explanation method.

\subsection{NLU examples}
\label{nlu_examples}

Natural-language understanding module is designed to guess an intent and extract relevant parameters/entities from a user query. Queries can be specified in a open format. Here are examples of NLU for three intents.

\begin{figure*}[b]
    \includegraphics[width=0.49\linewidth]{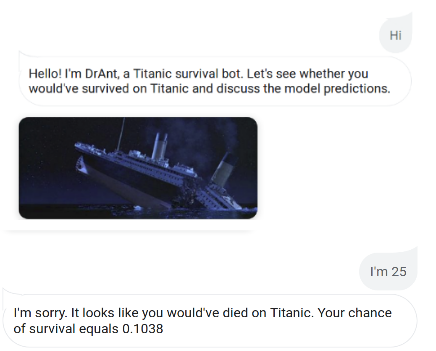}
    \includegraphics[width=0.49\linewidth]{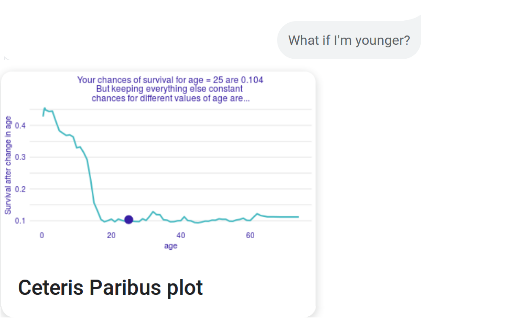}
    \caption{An example conversation. Explainee's queries in the grey boxes. }
    \label{example_conversation}
\end{figure*}

\noindent
    \textit{\textbf{Query}: What If I had been older?} \\
          \textit{\textbf{Intent:} ceteris\_paribus} \\
          \textit{\textbf{Entities:} [variable: age]} \\ \\
    \textit{\textbf{Query} I'm 20 year old woman} \\ 
          \textit{\textbf{Intent:} multi\_slot\_filling} \\
          \textit{\textbf{Entities:} [age: 20, gender: female]} \\ \\
    \textit{\textbf{Query:} Which feature is the most important?} \\
          \textit{\textbf{Intent:} break\_down} \\
          \textit{\textbf{Entities: []}} \\
\subsection{Example dialogue}
\begin{figure}
    \includegraphics[width=1\linewidth]{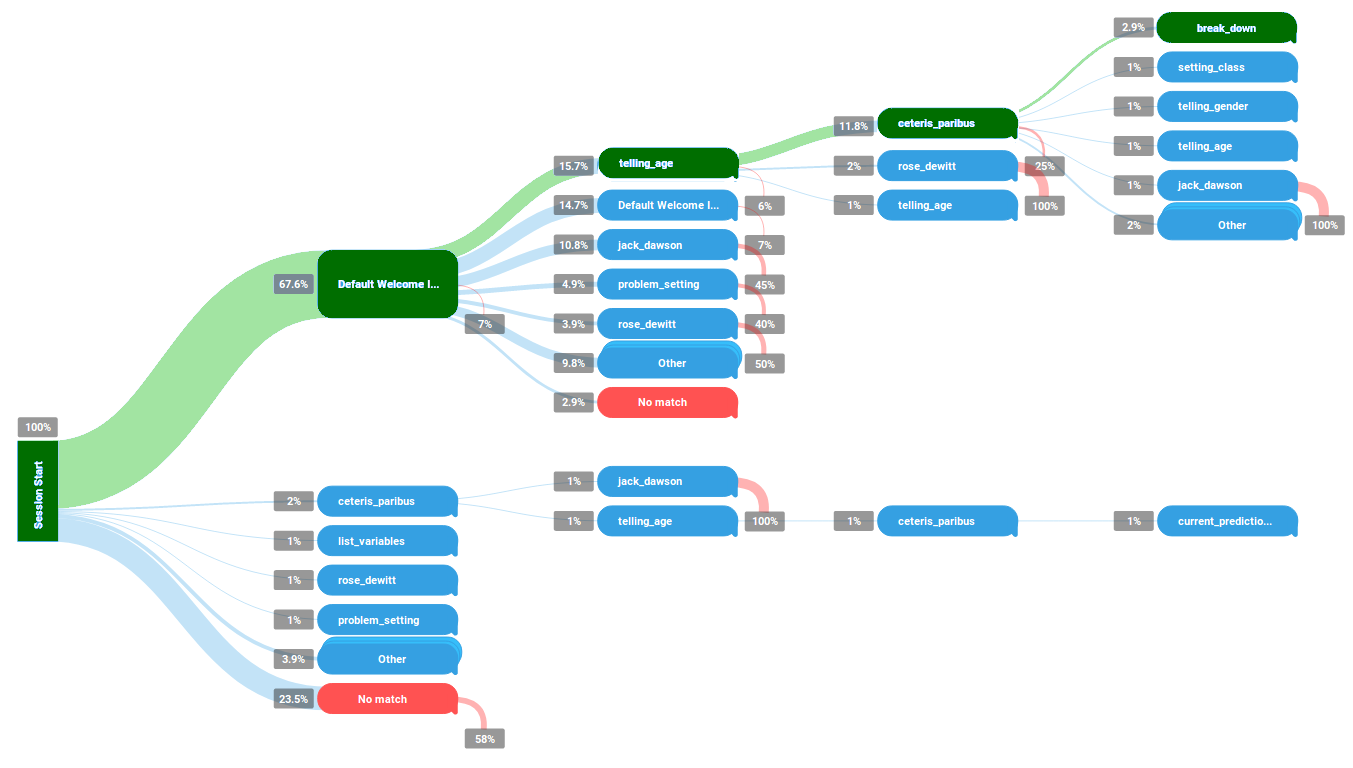}
    \caption{Screenshot from the \texttt{Dialogflow Analytics}. This flow chart demonstrates the results of the NLU module on a sample of collected dialogues. Example conversation from Figure~\ref{example_conversation} contributes to the topmost (green) path. Each box corresponds to a classified intention of the query, e.g. \textit{telling\_age} or \textit{ceteris\_paribus}.}
    \label{dialogflow}
\end{figure}

An excerpt from an example conversation is presented in  Figure~\ref{example_conversation}. The corresponding intent classification flow is highlighted in Figure~\ref{dialogflow}.

\section{Results}
\label{results_section}
The initial subset of the collected dialogues is used to improve the NLU module of the dialogue agent. As a next step, we conduct an experiment by sharing the chatbot in the Data Science community and analyzing the collected dialogues. 

\subsection{Experiment setup}
\begin{figure}
\centering
    \includegraphics[width=0.6\linewidth]{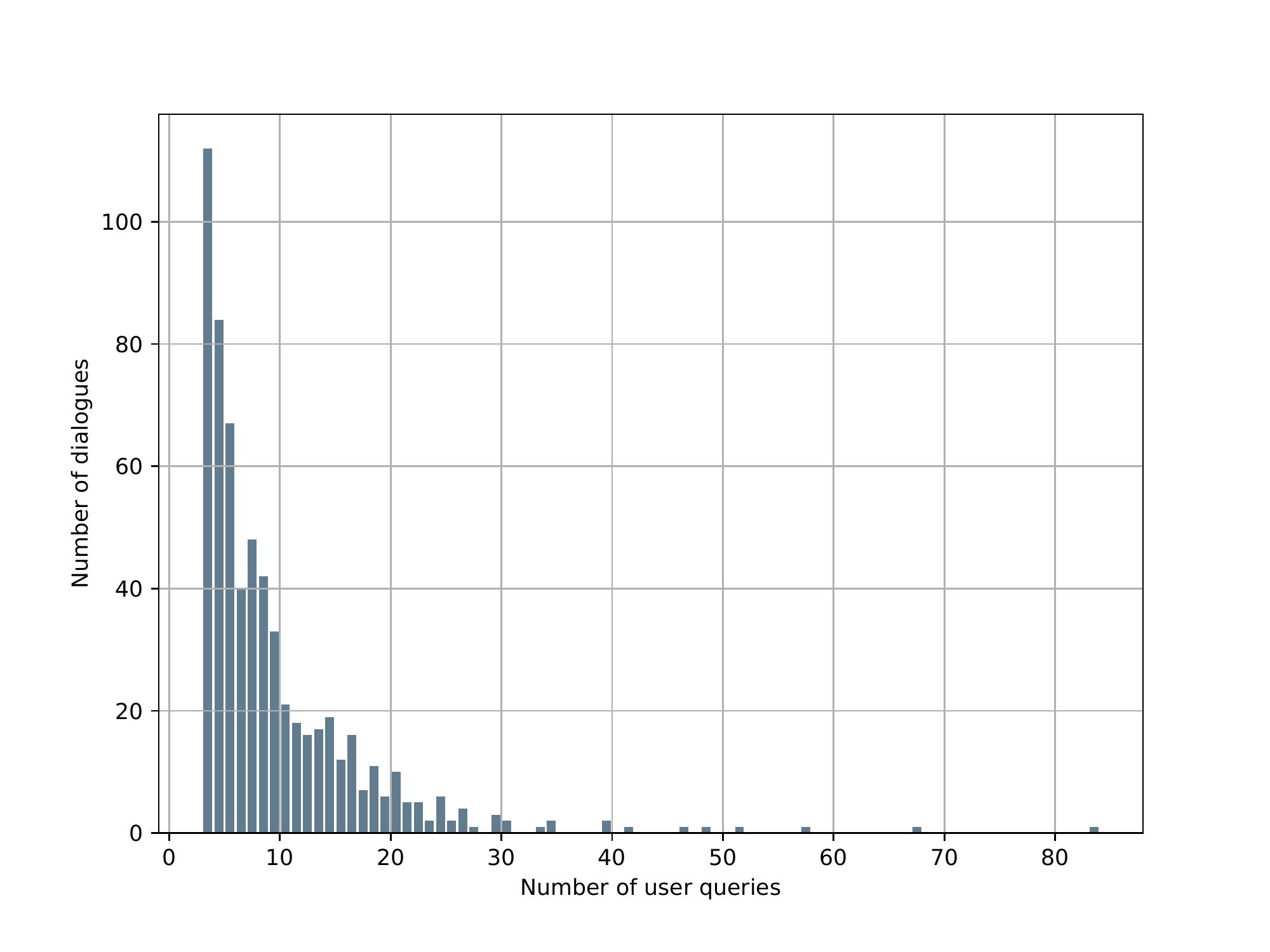}
    \caption{Histogram of conversations length (number of user queries), after filtering out conversations shorter than 3 queries. As expected, most conversations were short. However, there were also dialogues of over 20 queries.}
    \label{histogram_dialogues}
\end{figure}

For this experiment, we work on data collected throughout 2 weeks. This is a subset of all collected dialogues, separate from the data used to train the NLU module. Narrowing the time scope of the experiment allows to describe the audience and ensure the coherence of the data. As a next step, we filter out conversations with totally irrelevant content and those with less than 3 user queries. Finally, we obtain 621 dialogues consisting of 5675 user queries in total. The average length equals 9.14, maximum 83 and median 7 queries.
We see the histogram of conversations length in Figure~\ref{histogram_dialogues}.
Note that by conversation length we mean the number of user queries which is equal to the number of turns in the dialogue (user query, chatbot response).

The audience acquisition comes mostly from R and Data Science community. Users are instructed to explore the model and its explanations individually. However, they might come across a demonstration of the chatbot's capabilities potentially introducing a source of bias.

We describe the results of the study in the section~\ref{query_types} and we share the statistical details about the experiment audience in the section~\ref{population_statistics}.

\subsection{Query types}
\label{query_types}
We analyze the content of the dialogues. Similar user queries, when different only in the formulation, are manually grouped together. For each category, we calculate the number of conversations with at least one query of this type. Numbers of occurrences are presented in Table~\ref{queries_count}. 

Note that users were not prompted or hinted to ask any of these with an exception of the \textit{``what do you know about me''} question. Moreover, the taxonomy defined here is independent of the intents recognized by the NLU module and is defined based on collected dialogues.

Here is the list of the query types ordered decreasingly by the number of conversation they occur in.

\begin{enumerate}
    \item \textbf{why} -- general explanation queries, typical examples of such are:
        \begin{itemize}
            \item ``why?''
            \item ``explain it to me''
            \item ``how was this calculated?''
            \item ``why is my chance so low?''
        \end{itemize}
    \item \textbf{what-if} -- alternative scenario queries. Frequent examples: \textit{what if I'm older?}, \textit{what if I travelled in the 1st class?}. Rarely, we see multi-variable questions such as: \textit{What if I'm older and travel in a different class?}.
    \item \textbf{what do you know about me} -- this is the only query hinted to the user using the suggestion button. When the user inputs their data manually it usually serves to understand what is yet missing. However, in the scenario when the explainee impersonates a movie character it also aids understanding which information about the user is possessed by the system.
    \item \textbf{EDA} -- a general category on Exploratory Data Analysis. All questions related to data rather than the model fall into this category. For instance, \textit{feature distribution}, \textit{maximum values}, \textit{plot histogram for the variable v}, \textit{describe/summarize the data}, \textit{is dataset imbalanced}, \textit{how many women survived}, \textit{dataset size} etc.
    \item \textbf{feature importance} -- here we group all questions about the relevance, influence, importance or effect of the feature on the prediction. We see several subtypes of that query:
        \begin{itemize}
            \item \textit{Which are the most important variable(s)?}
            \item \textit{Does gender influence the survival chance?}
            \item \textbf{local importance} -- \textit{How does age influence my survival}, \textit{What makes me more likely to survive?}
            \item \textbf{global importance} -- \textit{How does age influence survival across all passengers?}
        \end{itemize}{}
    \item \textbf{how to improve} -- actionable queries for maximizing the prediction, e.g. \textit{what should I do to survive}, \textit{how can I increase my chances}.
    \item \textbf{class comparison} -- comparison of the predictions across different values of the categorical variable. It might be seen as a variant of the \textit{what-if} question. Examples: \textit{which class has the highest survival chance}, \textit{are men more likely to die than women}.
    \item \textbf{who has the best score} -- here, we ask about the observations that maximize/minimize the prediction. Examples: \textit{who survived/died}, \textit{who is most likely to survive}. It is similar to \textit{how to improve} question, but rather on a per example basis.
    \item \textbf{model-related} -- these are the queries related directly to the model, rather than its predictions. We see questions about the algorithm and the code. We also see users asking about metrics (accuracy, AUC), confusion matrix and confidence. However, these are observed just a few times.
    \item \textbf{contrastive} -- question about why predictions for two observations are different. We see it very rarely. However, more often we observe the implicit comparison as a follow-up question -- for instance, \textit{what about other passengers}, \textit{what about Jack}.
    \item \textbf{plot interaction} -- follow-up queries to interact with the displayed visual content. Not observed.
    \item \textbf{similar observations} -- queries regarding ``neighbouring'' observations. For instance, \textit{what about people similar to me}. Not observed.
\end{enumerate}

{\renewcommand{\arraystretch}{1.1} 
\begin{table}[]
\begin{center}
    \caption{Results of the analysis for 621 conversations in the experiment. The second column presents the number of conversations with at least one query of a given type. A single dialogue might contain multiple or none of these queries.}

    \begin{tabular}{lc}
\hline
\textbf{Query type}            & \textbf{Dialogues count} \\ \hline
why                       & 73             \\ 
what-if                   & 72             \\ 
what do you know about me & 57             \\ 
EDA                       & 54             \\ 
feature importance        & 31             \\ 
how to improve            & 24             \\ 
class comparison          & 22             \\ 
who has the best score    & 20             \\ 
model-related             & 14             \\ 
contrastive               & 1              \\ 
plot interaction          & 0              \\ 
similar observations      & 0              \\ \hline
\textbf{Number of all analyzed dialogues}            & \textbf{621}   \\ \hline
    \label{queries_count}

\end{tabular}
\end{center}{}
\vspace{-1cm}
\end{table}
}

We also see users creating alternative scenarios and comparing predictions for different observations manually, i.e. asking for prediction multiple times with different passenger information. Additionally, we observe explainees asking about other sensitive features, that are not included in the model, e.g. nationality, race or income. However, some of these, e.g. income, are strongly correlated with class and fare.

\subsection{Statistics of surveyed sample}
\label{population_statistics}
\begin{figure}[t!]
    \centering
    \includegraphics[width=0.53\linewidth]{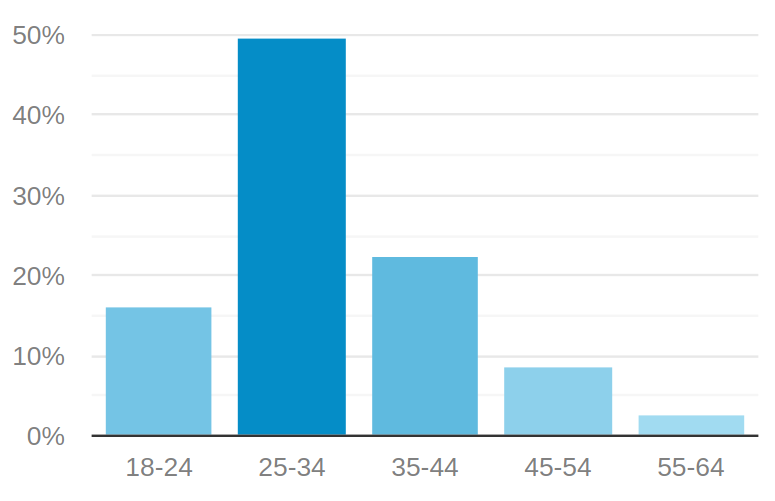}
    \includegraphics[width=0.37\linewidth]{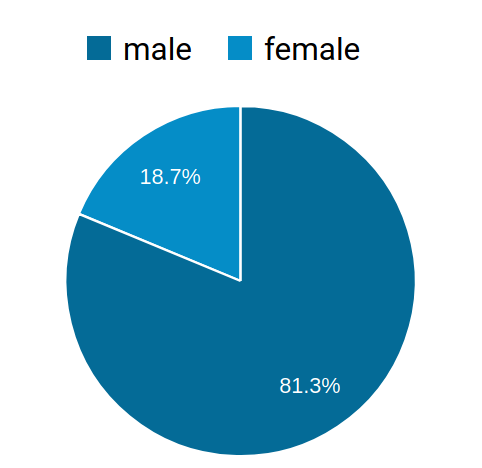}
    \caption{Demographic statistics for age (left) and gender (right) of the studied group registered by Google Analytics.}
    \vspace{-0.5cm}
    \label{audience_statistics}
\end{figure}

We use Google Analytics to get insights into the audience of the experiment. Users are distributed across 59 countries with the top five (Poland, United States, United Kingdom, Germany and India, in this order) accounting for 63\% of the users.
Figure~\ref{audience_statistics} presents demographics data on the subset of the audience (53\%) for which this information is available.

\section{Conclusions and Future Work}

Depending on the area of application, different needs are linked with the concept of interpretability \cite{Lipton2016,tomsett2018interpretable}. And even for a single area of application, different actors may have different needs related to model interpretability \cite{arya2019explanation}. 

In this paper, we presented a novel application of the dialogue system for conversational explanations of a predictive model.  Detailed contributions are following (1) we presented a process based on a dialogue system allowing for effective collection of user expectations related to model interpretation, (2) we presented a xai-bot implementation for a binary classification model for Titanic data, (3) we conducted an analysis of the collected dialogues.

We conduct this experiment on the survival model for Titanic. However, our prior goal of this work is to understand user needs related to the model explanation, rather than improve this specific implementation. The knowledge we gain from this experiment will aid in designing the explanations for various models trained on tabular data. One example might be survival models for COVID-19 which are currently under large interest.

Conversational agent proved to work as a tool to explore and extract user needs related to the use of the Machine Learning models. This method allowed us to validate hypotheses and gather requirements for the XAI system on the example from the experiment. In this analysis, we identified several frequent patterns among user queries. 

Conversational agent is also a promising, novel approach to XAI as a model-human interface. Users were given a tool for the interactive explanation of the model's predictions. In the future, such systems might be useful in bridging the gap between automated systems and their end users. An interesting and natural extension of this work would be to compare user queries for different explainee's groups in the system, e.g. model creators, operators, examiners and decision-subjects. In particular, it would be interesting to collect needs from explainees with no domain knowledge in Machine Learning. Similarly, it is interesting to take advantage of the process introduced in this work to compare user needs across various areas of applications, e.g. legal, medical and financial. Additionally, based on the analysis of the collected dialogues we see two related areas that would benefit from the conversational human-model interaction -- \emph{Exploratory Data Analysis} and \emph{model fairness} based on the queries about the sensitive and bias-prone features.

\section*{Acknowledgments}
We would like to thank 3 anonymous reviewers for their insightful comments and suggestions. 
Micha{\l} Ku\'zba was financially supported by the ‘NCN Opus \\grant 2016/21/B/ST6/0217’.

\bibliographystyle{splncs04}
\bibliography{paper}

\end{document}